\documentclass[9pt,journal]{IEEEtran}

\usepackage[ruled, linesnumbered]{algorithm2e}

\usepackage{algorithmicx,algpseudocode} 
\usepackage{cite}
\usepackage{graphicx}
\usepackage{amsmath,amssymb,amsfonts,amsthm}

\usepackage[font=footnotesize,labelfont=bf]{caption}
\usepackage[font=scriptsize,labelfont=bf]{subcaption}
\usepackage{subcaption}

\usepackage{amsmath}
\usepackage{comment}

\usepackage{verbatim,url}
\usepackage{float}
\usepackage{color}
\usepackage{graphicx}
\usepackage{amsmath,amsfonts,amssymb}
\usepackage{multirow}
\usepackage{hhline}
\usepackage{amsthm}
\usepackage{verbatim, url}

\newcounter{theorem}
\newcounter{def}
\newcounter{mylea}
\newcounter{corollary}
\usepackage{fancyhdr}

\newcommand{\mbf}{ }
\newcommand{\mvect}{ \vec }

\newcommand{\at}[2]{\left.#1\right|_{#2}}

\makeatletter
\renewcommand\@makefntext[1]{%
\noindent\makebox[0em][r]{\@makefnmark}#1}
\makeatother

\begin{document} 

\title{\bf \huge An Algorithmic Framework for Efficient Large-Scale Circuit Simulation  Using  Exponential Integrators}
\author{ 
	{Hao~Zhuang$^1$, Wenjian Yu$^2$, 
	Ilgweon Kang$^1$, Xinan Wang$^1$, and  Chung-Kuan~Cheng$^1$\\} 
{$^1$Department of Computer Science \& Engineering, University of California, San Diego, CA, USA\\
$^2$Department of Computer Science \& Technology, Tsinghua University, Beijing, China\\       hao.zhuang@cs.ucsd.edu,~yu-wj@tsinghua.edu.cn,~\{igkang,~xinan,~ckcheng\}@ucsd.edu\\}
\thanks{Permission to make digital or hard copies of all or part of this work for personal or classroom use is granted without fee provided that copies are not made or distributed for profit or commercial advantage and that copies bear this notice and the full citation on the first page. Copyrights for components of this work owned by others than ACM must be honored. Abstracting with credit is permitted. To copy otherwise, or republish, to post on servers or to redistribute to lists, requires prior specific permission and/or a fee. Request permissions from Permissions@acm.org.}
\thanks{\textit{DAC '15}, June 07 - 11, 2015, San Francisco, CA, USA} 
\thanks{Copyright 2015 ACM 978-1-4503-3520-1/15/06$\$15.00$} 
\thanks{http://dx.doi.org/10.1145/2744769.2744793}
}

\IEEEoverridecommandlockouts
\IEEEaftertitletext{\vspace{-2\baselineskip}}

\maketitle 

\thispagestyle{fancy}

\begin{abstract} 
We propose an efficient algorithmic framework for time-domain circuit simulation using exponential integrators. This work addresses several critical issues exposed by previous matrix exponential based circuit simulation research, and makes it capable of simulating stiff nonlinear circuit system at a large scale. 
In this framework, the system's nonlinearity is treated with exponential Rosenbrock-Euler formulation. 
The matrix exponential and vector product is computed using invert Krylov subspace method. 
Our proposed method has several distinguished advantages over conventional formulations (e.g., the well-known backward Euler with Newton-Raphson method). 
The matrix factorization is performed only for the  conductance/resistance matrix $\mbf G$, without being performed for the combinations of the capacitance/inductance matrix $\mbf C$ and matrix $\mbf G$, which are used in traditional implicit formulations. 
Furthermore, due to the explicit nature of our formulation,
we do not need to repeat LU decompositions when adjusting the length of time steps for error controls. Our algorithm is better suited to solving tightly coupled post-layout circuits in the pursuit for full-chip simulation.
Our experimental results validate the advantages of our framework.

\end{abstract}
\vspace{-0.15in}

\section*{ \bf{Categories and Subject Descriptors}} 
\vspace{-0.05in}
\noindent B.7.2 [{\bf Integrated Circuits}]: Design Aids - {\emph {simulation}}
\vspace{-0.15in}

\section*{\bf{General Terms}}
\vspace{-0.05in}
\noindent Algorithms, Design, Performance
\vspace{-0.15in}

\section*{\bf{Keywords}}  
\vspace{-0.03in}
\noindent Circuit Simulation, Transient Simulation, Exponential Integrators 
 \vspace{-0.1in}

\section{Introduction}
\label{sec_intro}
SPICE-like transistor-level circuit simulation \cite{Nagel75, Najm10_book, Leon75} of integrated circuits is indispensable during the cycle of very large scale integration (VLSI) designs. It is crucial to cost-efficient production of VLSI chips. 
Smaller process geometries, larger designs as well as tighter design margins translate to the need for more accurate large-scale circuit simulation, e.g., post-layout simulations with more detailed parasitic couplings \cite{SILCA, Joel01_ICCAD, Zhu07_TCAD}.
 With advancing technologies, three dimensional IC structures, and increasing complexities of system designs,  
the numbers of chip components are easily more than millions \cite{ljw15_TCAD, ljw14_DAC},
which make simulation tasks 
very time-consuming. Finding effective algorithms still remains challenging and has always been an important research topic for several decades \cite{Nagel75, Leon75, Najm10_book}.


Time-domain circuit simulation algorithm relies on numerical integration to solve differential algebraic equations (DAE). To capture circuit's transient behaviors, numerical integration is computed step by step till the end of whole simulation time span. Conventional numerical integration formulations are forward Euler (FE), backward Euler (BE), Trapezoidal (TR), Gear's and multi-step methods \cite{Nagel75, Leon75, Najm10_book}. 
Despite explicit formulation, such as FE, avoids solving linear system, the time steps are usually restricted for ensuring numerical stability \cite{He12_TAP, Dong09, Leon75, Najm10_book}.
In general, VLSI circuits form stiff and nonlinear dynamical systems. Therefore, the implicit formulations, e.g., BE, 
are much more stable, robust and widely deployed in VLSI circuit simulators.
Based on the implicit formulation of such nonlinear system, Newton-Raphson (NR) iterations are typically adopted to obtain solutions. 
Each NR iteration needs to linearize system and then solves the linear system. During this process, direct  solvers \cite{Davis06, nicslu} have been favored and applied because of their robustness and ease of use. 
However, it is known that direct solvers, e.g., LU decomposition,
have super-linear computational complexities and very expensive to simulate large-scale and strongly coupled circuit systems (the cost of LU decomposition can approach the worst case $O(n^3)$ \cite{SILCA, Joel01_ICCAD}).
It is crucial to devise efficient algorithms to reduce the numbers of expensive LU operations and NR iterations in the  circuit simulation.
For example, leveraging the explicit formulations is a research direction \cite{Dong09, He12_TAP}.

Above formulations are all categorized into the low order approximation of DAE, which is with the order typically lower than ten \cite{Leon75}. 
The error of those formulations are proportional to the step size due to local truncation error (LTE).  Beyond those conventional low order schemes,  a high order matrix exponential based time integration kernel \cite{Leon75} has been recently triggering academic researchers' interests because of the progress of efficient matrix function computation using Krylov subspace methods 
\cite{Saad92, Hochbruck10_EXP, Hochbruck09_RB, Caliari09_ER}. 
In VLSI CAD and EDA research, this exponential integration kernel has been applied in time domain electromagnetic and technology semiconductor device simulation \cite{Chen12_ICCAD}, power distribution network analysis \cite{Zhuang13_ASICON, Zhuang14_DAC} as well as general circuit simulation \cite{Weng12_TCAD, Weng12_ICCAD}.
Those frameworks provide analytical solution for transient simulation of linear system, and better properties in numerical accuracy, stability and time step controlling, etc \cite{Leon75, Caliari09_ER,  Zhuang13_ASICON, Zhuang14_DAC}. 
However, one drawback of previous works for matrix exponential based circuit simulation \cite{Weng12_TCAD}
is the slow convergence rate for matrix exponential and vector product ({MEVP}) by using standard Krylov subspace \cite{Chen12_ICCAD, Weng12_TCAD}. 
In addition, the nonlinear formulation still requires NR iterations \cite{Chen12_ICCAD, Weng12_TCAD}, 
which does not fully utilize the explicit
nature from matrix exponential formulation \cite{Hochbruck10_EXP}.

 
In this work, two major processes are devised to address above challenges. First, we propose a new matrix exponential based framework using exponential Rosenbrock-Euler integration formulation \cite{Caliari09_ER, Hochbruck09_RB}, which well suits nonlinear dynamical differential equation problems and preserves explicit features. Then, we also design error control scheme for adaptive time stepping. 
Second, we deploy invert Krylov subspace method \cite{Zhuang14_DAC} to compute MEVP in an efficient manner.  
The {{major contributions}} are listed as follows. 
\begin{itemize}
\item By invert Krylov subspace strategy, we improve the convergence rate for the MEVP computation, the key operation within the matrix exponential based method.  
Besides, this approach also removes the regularization process \cite{Weng12_TCAD, Chen12_TCAD}, which would be impractical for large designs with singular matrix $\mbf C$. Therefore, this approach enables the application for the simulation of large stiff VLSI designs.
\item Proposed invert Krylov subspace strategy removes 
$\mbf C$ from matrix factorization processes. 
Building invert Krylov subspace bases only needs to factorize $\mbf G$, which is much sparser and simpler than $\mbf C$ in 
strongly coupled post-layout circuits. 
In contrast, conventional implicit methods require LU decomposition of the combinations of $\mbf G$ and $\mbf C$. 
Therefore, our method is better suited to handle those 
strongly coupled post-layout circuits 
in the pursuit for full-chip simulation. 
\item We devise  exponential Rosenbrock-Euler formulation for stable  explicit DAE solver, as well as  correction term  on top of that to further improve the accuracy. 
The stability of the explicit formulation is preserved by the high-order approximation of exponential operator \cite{Saad92, Hochbruck09_RB, Caliari09_ER}. 
Thus,
it takes only one LU decomposition per time step, while conventional methods, e.g., BE with NR (BENR), 
require at least two times of LU to verify the convergence.
\item Moreover, our method does not need to repeat LU when we adjust the length of time steps for error controls. It is because our  matrix exponential based explicit formulation and  (time-step) scaling-invariant  property of Krylov subspace \cite{Saad92, Weng12_TCAD}. 
On the contrary, the low order approximation schemes force time step embedded in the linear matrix and conduct matrix factorization. Once the time step is adjusted, LU is unavoidable in order to solve the new linear system. 
\item We test our framework with baseline BENR against our large-scale circuits. For some challenging test cases, our framework can achieve speedups by magnitude, and even manage to finish them when conventional counterpart fails. 
\end{itemize}


The remainder of this paper is organized as follows. Sec. \ref{sec_back} introduces the background of nonlinear circuit simulation. Sec. \ref{sec_nonlinear} presents our new circuit simulation framework that  utilizes exponential Rosenbrock-Euler method and related formulations.
Sec. \ref{sec_krylov} illustrates the MEVP computation uses invert Krylov subspace method. Sec. \ref{sec_result} shows experimental results and  validates our framework. Sec. \ref{sec_conclusion} concludes this paper.

 \vspace{-0.12in}
\section{Background}
\label{sec_back}
\begin{figure*}[t]
\centering
\includegraphics[trim = 0.6cm 18cm 6cm 0.3cm, clip,keepaspectratio,width=1.52\textwidth]{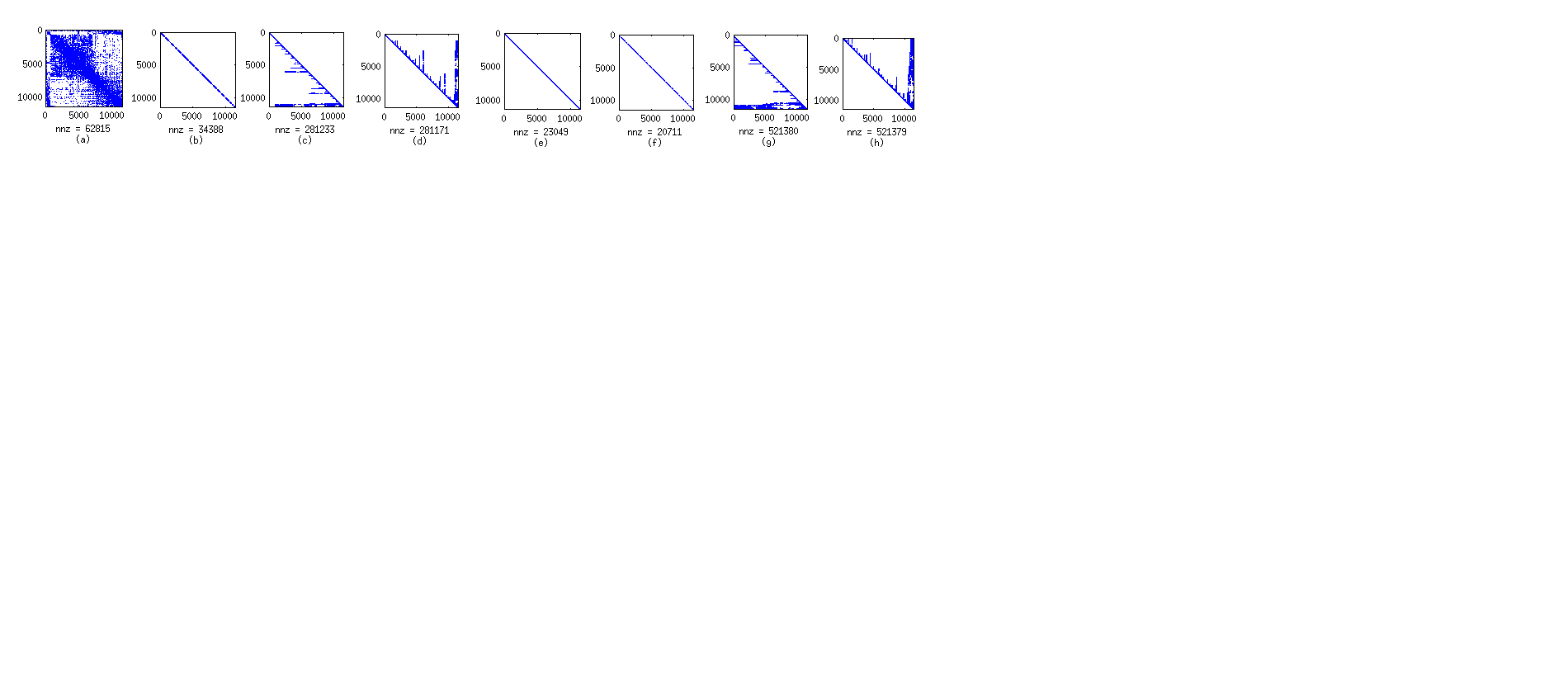}
\vspace{-0.7cm}
\caption{Visualization of post-extraction matrices' non-zero elements distributions from a design {\bf FreeCPU} \cite{Yu14_TCAD}, the sizes of matrix are ${11417 \times 11417}$, which are obtained from SPEF extracted by industrial tool \emph{Synopsys Star-RCXT}. $nnz$ is the number of non-zeros in the matrix. {(a) Extracted capacitance matrix $\mbf C$ (non-zero entries distribute widely in the matrix). (b)   Extracted conductance matrix $\mbf G$ (there are many off-diagonal non-zeros in the matrix, but the bandwidth is much smaller than $\mbf C$).
(c) Lower triangular matrix $\mbf L_C$ and 
(d) Upper triangular matrix $\mbf U_C$ of $LU\_decompose(\mbf C)$;}
(e) Lower triangular matrix $\mbf L_G$ and  
(f) Upper triangular matrix $\mbf U_G$ of $LU\_decompose(\mbf G)$;
(g) Lower triangular matrix $\mbf L_{\frac{\mbf C}{h} +\mbf G}$ and 
(h) Upper triangular matrix $\mbf U_{\frac{\mbf C}{h}+ \mbf G}$ of $LU\_decompose(\frac{\mbf C}{h}+ \mbf G)$.
The function of $LU\_decompose$ uses MATLAB2013a UMFPACK.   
$\mbf L_G$ and $\mbf U_G$ contain much smaller $nnz$ than $\mbf L_C$, $\mbf U_C$, $\mbf L_{\frac{\mbf C}{h}+\mbf G}$ and $\mbf U_{\frac{\mbf C}{h} + \mbf G}$. 
}
\label{fig_matrix_distribution}
\vspace{-0.3cm}
\end{figure*}


Given a circuit netlist and device models, time-domain simulation of general nonlinear electronic circuit is formulated as  
\begin{eqnarray}
\label{eq_dae}
\frac{d   \mvect q ( \mvect  x(t))}{dt} + \mvect  f( \mvect  x(t) ) =  \mbf B  \vec  u(t),
\end{eqnarray} 
where $\mvect x(t) \in \mathbb{R}^{n \times 1}$ denotes nodal voltages and branch currents and $n$ is the length of vector. 
$\mvect q \in \mathbb{R}^{n \times 1}$ and $\mvect f \in \mathbb{R}^{n \times 1}$ represent the charge/flux and current/voltage terms, respectively. 
$\mvect u(t)$ is a vector representing all the external excitations at time $t$; $B$ is an incident matrix that inserts those signals to the system. 
\vspace{-0.15in}

\subsection{Circuit Simulation Using Low Order Integration Schemes}
\label{sec_intro_conv}

The conventional numerical approaches discretized Eq. (\ref{eq_dae}) with time step size $h_k = t_{k+1} - t_{k}$.
To compute the solution $\mvect x_{k+1}$ at $t_{k+1}$,  we start from an initial solution $\mvect x_k$. The well-known BE, which is a stable implicit integration scheme, leads to 
\begin{eqnarray} 
  \frac{     \mvect  q(\mvect x_{k+1} )  -  \mvect  q(\mvect x_{k})}{h_k}   +
    \mvect f(\mvect x_{k+1} ) 
    = \mbf B \mvect u(t_{k+1})  
    \label{eq_be}  
\end{eqnarray} 
Usually NR is a widely adopted method of solving this implicit formulation Eq. (\ref{eq_be}). 
Each NR iteration, direct solver (e.g., LU decomposition) is applied to solve 
\begin{eqnarray}
\frac{\partial {\mvect T ({\mvect x^i})}}{\partial {\mvect x}} (\mvect x^{i+1} -\mvect x^{i}) = - \mvect T(\mvect x^{i})
\label{eq_NR}
\end{eqnarray}
until it is convergent, which means the difference of  the solution from $i$-th iteration $\mvect x^{i}$ and $\mvect x^{i+1}$ is ``small enough''. 
$\mvect T(\mvect x )=\frac{\mvect  q(\mvect x )  -  \mvect  q(\mvect x_{k})}{h_k}  +    \mvect f(\mvect x )  - \mbf B \mvect u(t_{k}+h_k) $
corresponds to Eq. (\ref{eq_be}). Then we have  $\frac{\partial {\mvect T ({\mvect x^i})}}{\partial {\mvect x}} = \left( \frac{C(\mvect x^i)}{h}+G(\mvect x^i) \right) \in \mathbb{R}^{n \times n}$ is the Jacobian matrix of $\mvect T$, where   $ \mbf C(\mvect x^i) \in \mathbb{R}^{n \times n}$ is a matrix of capacitances and inductances linearized at $\mvect x^i$; 
$\mbf G(\mvect x^i) \in \mathbb{R}^{n \times n}$ represents the linearized  resistances and conductances as well as the incidence between voltages and currents. 
This approach usually serves a \emph{de facto} method in industrial tools and academic research, which is called as \emph{BENR} in this paper. To solve Eq. (\ref{eq_dae}), we can also deploy other low order implicit discretized forms, such as TR and Gear's, with NR.

Due to the low order implicit integration scheme, Eq. (\ref{eq_NR}) embeds time step $h_k$ in $\mvect T$ and $\frac{\partial {\mvect T ({\mvect x^i})}}{\partial {\mvect x}}$ 
 (following the practice of SPICE). 
If the estimated LTE violates numerical error budget,  $h_k$ will be reduced. Then new NR iterations for $\mvect x_{k+1}$ are re-launched with the updated $h_k$.
Besides, $\mbf C$ always exists in Jacobian matrix   $\frac{\partial {\mvect T ({\mvect x^i})}}{\partial {\mvect x}}$.
Large volume of non-zeros of $\mbf C$ are introduced  from the post-layout parasitics extraction \cite{Yu14_TCAD, Zhuang2012, Yu2013}, resulting to huge computational challenges for the capability of numerical integration algorithms \cite{Zhu07_TCAD} and model order reductions \cite{SparseRC}.
In addition, the non-zero fill-ins of off-diagonal terms in $\mbf C$ and $\mbf G$ are usually mutually exclusive in VLSI circuits.
For example, the matrices in Fig. \ref{fig_matrix_distribution} are from post-extraction of a design $\bf FreeCPU$ \cite{Yu14_TCAD}.
We notice that the distribution of non-zeros in $C$ (Fig. \ref{fig_matrix_distribution}(a)) is much more complicated than $G$ (Fig. \ref{fig_matrix_distribution}(b)). After LU decompositions of those matrices, the factorized matrices Fig. \ref{fig_matrix_distribution}(c) and Fig.  \ref{fig_matrix_distribution}(d) of $C$, Fig. \ref{fig_matrix_distribution}(g) and Fig. \ref{fig_matrix_distribution}(h) of $\left ( \frac{\mbf C}{h}+G\right)$ also contain much larger $nnz$ (number of non-zeros) than Fig. \ref{fig_matrix_distribution}(e) and Fig. \ref{fig_matrix_distribution}(f) of $G$. 

To address above challenges, we use exponential integrators \cite{Hochbruck10_EXP} and architect a framework with stable explicit formulation.  
The next sub-section briefs previous progresses of circuit simulation using related exponential integration methods and the major problems, which keep them from the large-scale circuit simulation. 
\vspace{-0.15in}

\subsection{Circuit Simulation Using Exponential Integrators}

Recently, many works are proposed to solve time-domain simulation problem using  exponential integrators \cite{Weng12_TCAD, Chen12_ICCAD, Weng12_ICCAD, Zhuang13_ASICON, Zhuang14_DAC}.
Chen  \emph{et al.}
\cite{Chen12_ICCAD} and Weng \emph{et al.} \cite{Weng12_TCAD} adopted a technique developed in \cite{Nie06}, which decouples the linear and nonlinear terms by approximating the integrand in Eq. (\ref{eq_mexp}) with Lagrange polynomial.
\begin{eqnarray}
\label{eq_mexp}
\mvect x_{k+1} &=& e^{h_k \mbf J (\mvect x_k)} \mvect x_k +\\
& & 
\int_0^{h_k} e^{(h_k -\tau) 
\mbf J (\mvect x_k)}\mbf C^{-1}(\mvect x_k) 
\left (\mvect F (\mvect x_k+\tau) + 
  \mbf B \mvect u(t_k+\tau)   \right )d\tau \nonumber
\end{eqnarray}
where $\mbf J (\mvect x ) = -\mbf C^{-1}(\mvect x )\mbf G(\mvect x)$, and  $\mvect F(x)$ collects the nonlinear dynamics at $\mvect x$ from transistor models, such as BSIM3. 
The second-order approximation yields $\mvect x_{k+1}$ as the solution of a nonlinear equation, which again
can be solved by NR using Eq. (\ref{eq_NR}). 
The Jacobian matrix in Eq. (\ref{eq_NR}) is
  $\mbf C(\mvect x^i) + \frac{h}{2} \frac{\partial \mvect F(\mvect x^i)}{\partial \mvect x}$.
More details can be found in \cite{Weng12_ICCAD, Chen12_ICCAD}. 
Therefore, it is similar to conventional method in Sec. \ref{sec_intro_conv}.
During NR iterations, the matrix factorization involves $\mbf C$ in the Jacobian matrix.
Then it will suffer from the runtime degradation when 
$\mbf C$ is relatively large and complicated.

Another problem comes from the key operation of $e^{\mbf J}\mvect x$, the matrix exponential-and-vector product (MEVP), where $\mbf J$ is a $n \times n$ matrix.
MEVP is computed by \emph{standard Krylov subspace} \cite{Saad92, Weng12_TCAD, Weng12_ICCAD, Chen12_ICCAD}. 
Such Krylov subspace 
is constructed as
 \begin{eqnarray}
 {\rm K_m}(\mbf J, \mvect x) := \text{span}
{\{ \mvect v_1,   \mvect v_2,
\cdots,  \mvect v_{m}}\} =
 \text{span}
{\{ \mvect x, \mbf J \mvect x,
\cdots, \mbf  J^{m-1}\mvect x}\}.
\label{eq_krylov}
\end{eqnarray}
An $n \times m$ orthonormal basis and an $(m+1) \times m$ upper Hessenberg matrix $\mbf H$ for the Krylov subspace ${\rm K_m}$ are first constructed by Arnoldi process \cite{Chen12_ICCAD, Weng12_TCAD, Weng12_ICCAD}.
Then MEVP is computed via 
\begin{eqnarray}
e^{h\mbf  J} \mvect v \approx \lVert \mvect v \rVert \mbf V_m e^{h\mbf {H}_m} \mvect e_1
\label{mevp}
\end{eqnarray}
where $\mbf H_m = \mbf H(1:m,1:m)$, $\mbf e_1 \in \mathbb{R}^{m \times 1}$ and $\mbf e_1$ is the first unit vector.
When $\mbf A$ is mildly stiff, $\mbf H_m$ is usually much smaller than original $\mbf A$ \cite{Weng12_TCAD}.
However, when the values in $\mbf C$ vary in magnitudes caused by stiff circuit system, standard Krylov subspace demands large dimension of subspace to approximate MEVP, then degrades performance of matrix exponential-based circuit simulation. Such phenomenon has been  observed in power distribution network simulation using matrix exponential framework \cite{Zhuang13_ASICON, Zhuang14_DAC}.  
Besides, $\mbf C$ should not be singular during the process of standard Krylov subspace strategy. Otherwise, related regularization process  \cite{Weng12_TCAD, Chen12_TCAD} is required, which is time-consuming for large-scale designs. 




\section{Circuit Simulation Using Exponential Integrators Based on Rosenbrock-Euler Formulation}
\label{sec_nonlinear}
In contrast to standard low order and previous matrix exponential integration schemes, our framework adopts exponential Rosenbrock-Euler method, and leads to a stable  explicit method, which does not require NR iterations \cite{Hochbruck09_RB}.
To simplify the derivations from exponential Rosenbrock-Euler
to SPICE-like formulation, we consider a non-autonomous ordinary differential equation (ODE) system,
\begin{eqnarray}
\frac{d\mvect x(t)}{dt} = \mvect  g(\mvect x, \mvect u,t). 
\end{eqnarray}
Exponential Rosenbrock-Euler method is derived to  compute $\mvect x_{k+1}$ with step size $h_k$ as follows,
\begin{eqnarray}
\mvect x_{k+1} =  \mvect  x_{k} + h_k \phi_1(h_k \mbf J_k) \mvect g(\mvect x_k,\mvect u, t_k)+h^2_k\phi_2(h_k\mbf J_k)\mvect b_k
\label{eq_rbeuler}
\end{eqnarray}
where $\mbf J_k  $ denotes the 
Jacobian matrix of $\mvect g$,
and $
\mvect b_k = \frac{\partial  \mvect g}{\partial  t} (\mvect x_k, \mvect u, t_k) 
$.
\cite{Hochbruck10_EXP, Hochbruck09_RB, Caliari09_ER}.  
$\phi_{i}(z)$ describes ODE's time evolution, where $\phi_{i}(z) =  \int_{0}^1 e^{z(1-s)}\frac{s^{i-1}}{(i-1)!}ds  ~~(i \ge 1)  
$.
We have $\phi_{0}(h_k \mbf J_k) = \mbf e^{h_k \mbf J_k}$ and $ \mbf I_n \in \mathbb{R}^{n \times n}  $  is the identity matrix, then 
\begin{eqnarray}
\phi_1(h_k \mbf J_k) = \frac{e^{h_k \mbf J_k } - \mbf I_n}{h_k \mbf J_k} , ~~~
\phi_2(h_k \mbf J_k) = \frac{ e^{h_k \mbf J_k} - \mbf I_n}{h^2_k \mbf J^2_k} - \frac{I_n}{h_k  \mbf J_k}. 
\label{eq_phi} 
\end{eqnarray}
More details can be found in the works of Hochbruck and Ostermann \cite{Hochbruck10_EXP, Hochbruck09_RB}. 
The advantage of this formulation is the explicit nature and the superior stable region (in the entire complex plane) than the class of low order integrations schemes.
Therefore $\phi_i$ functions permit a large value for the step size $h_k$ with guaranteed stability \cite{Hochbruck10_EXP, Hochbruck09_RB, Caliari09_ER}.
 
To utilize above formulation in SPICE-like circuit simulation, we apply the chain rule to Eq. (\ref{eq_dae}), 
\begin{eqnarray}
\frac{d \mvect q(\mvect x(t))}{d \mvect x}\cdot  \frac{d \mvect x(t)}{dt}
=  \mbf B \mvect u(t) -f(\mvect x).
\label{eq_chain}
\end{eqnarray}
Eq. (\ref{eq_chain}) is linearized at the state $\mvect x_k$,
when $\frac{d \mvect q(\mvect x_k)}{d \mvect x} 
= \mbf C{(\mvect x_k)} = C_k$. We obtain the equivalent format for the non-autonomous dynamical system 
\begin{eqnarray}
\frac{d\mvect x(t)}{dt}  = 
\mvect  g(\mvect x, \mvect u, t)=  \mbf J_k  \mvect x(t) 
  + \mbf C^{-1}_k 
  \left(  \mvect F(\mvect x(t)) 
   +  
   \mbf B \mvect u(t)   \right)
\end{eqnarray}
where $\mbf J_k = -\mbf C^{-1}_k \mbf G_k$ and $\mbf G_k = \mbf G \left(\mvect x_k \right)$. 
The circuit system response is computed by high order function $h_k \phi_1(h_k \mbf J_k)\mvect  g_k$,  where
\begin{eqnarray}
\mvect g_k = \mbf J_k\mvect x_k + \mbf C^{-1}_k  \left(  \mvect F_k
 +  \mbf B \mvect u(t_k) \right),
\label{eq_yk}
\end{eqnarray}  
and $\mvect F_k = F \left(\mvect x_k \right)$.
Furthermore, we assume $\mvect u(t)$ is a piecewise-linear function for $t \in [t_k,t_{k+1}]$  in VLSI designs, so the contribution from external excitations can also be captured by $h_k^2 \phi_2(h_k \mbf J_k)\mvect b_k$ \cite{Hochbruck10_EXP, Hochbruck09_RB, Caliari09_ER}, where 
\begin{eqnarray}
\label{eq_two_vec1}
\mvect b_k =   \mbf C^{-1}_k \mbf B  \frac{ \mvect u(t_{k+1}) - \mvect u(t_k)}{h_k}
\end{eqnarray}
The next time step solution $\mvect x_{k+1}$ with $h_k$ is 
\begin{eqnarray}
 \mvect x_{k+1}(h_k) = \mvect x_k + h_k \phi_1 (h_k \mbf J_k ) \mvect g_{k} + h_k^2 \phi_{2} (h_k\mbf J_k) \mvect b_k. 
 \label{eq_rbe}
\end{eqnarray}
Note the denominator of $\phi_i$ in Eq. (\ref{eq_phi}) always cancels out  $\mbf C^{-1}_k$ in Eq. (\ref{eq_yk}) and Eq. (\ref{eq_two_vec1}). 
Hence, we avoid the matrix factorization of $\mbf C_k$. 
\vspace{-0.1in}  
  
\subsection{Local Nonlinear Error Estimator}
The error estimator is an important ingredient of adaptive time-stepping for the nonlinear systems. Based on the notion proposed by Caliari and Ostermann \cite{Caliari09_ER}, we devise a formula to estimate local nonlinear truncation error to control the step size,
\begin{eqnarray}
  \label{eqn_expRB2_err}
     e_{rr} (\mvect x_{k+1},\mvect x_k)   
    =  h_k\phi_1(h_k\mbf J_k) \mbf C^{-1}_k \Delta \mvect F_k =  - \left( e^{h_k \mbf J_k} -\mbf I_n \right) \mbf G^{-1}_k \Delta \mvect F_k,  
\end{eqnarray}
where $\Delta \mvect F_k =  \mvect F_{k+1} - \mvect F_{k}$.
Intuitively, Eq. (\ref{eqn_expRB2_err}) represents the response changes inside the nonlinear system along time evolutions. 
When the strong nonlinear phenomenon is detected, 
the absolute value of $e_{rr}$ becomes large and shrinks the  step size to obtain an accurate enough solution. 

\subsection{Modified Correction Term for Exponential Rosenbrock-Euler Method}
\label{sec_nonlinear_correct}
Furthermore, we can reuse the intermediate results $\Delta \mvect F_k$ (after device evaluations at $\mvect x_{k+1}$) of Eq. (\ref{eqn_expRB2_err}) to   improve the accuracy of solution \cite{Caliari09_ER}
The correction strategy is designed by utilizing $\phi_2$
\cite{Caliari09_ER},  
\begin{eqnarray}
\mvect D_k = \gamma h_k \phi_2(h_k\mbf J_k)  \mbf C^{-1}_k  \Delta \mvect F_k,  ~~~
\end{eqnarray}
where $\gamma$ is constant and has several choices \cite{Caliari09_ER, Hochbruck10_EXP}.  
Within the time step, by adding this correction term, the corrected solution  $\mvect x_{k+1,c}$ is  
\begin{eqnarray}
\mvect x_{k+1, c} =\mvect  x_{k+1}  -  \mvect D_k.
 \label{eq_ferc}
\end{eqnarray}
In order to obtain such more accurate $\mvect x_{k+1,c}$ 
by reusing $\Delta \mvect F_k$, 
we need extra computations, including Krylov subspace generations.  



\section{MEVP Using Invert Krylov Subspace}
\label{sec_krylov}
Krylov subspace method enables the time-domain circuit simulation algorithm to use exponential integrators.
The key of efficient Krylov subspace based matrix function evaluation is keeping the size of the Krylov basis $m$ small so that calculation of Eq. (\ref{mevp})
is cheap.
 However, using standard Krylov subspace for MEVP 
$e^{ \mbf  J}\mvect v$ 
with stiff $\mbf J ={-\mbf C^{-1} \mbf G} $ is not efficient enough  due to 
the demand of large dimensional subspace.  
Our interpretation of the inefficiency 
is standard  Krylov subspace method tends to oversample eigenvalues with large magnitudes in the spectrum of $\mbf J$, which are not crucial players to define dynamical behaviors  \cite{Zhuang13_ASICON, Zhuang14_DAC}. 

Another drawback of standard Krylov subspace comes from matrix factorization of $\mbf C$ for subspace generations.  
To generate basis $\mbf v_{i+1}$, we need to solve $ \mbf C \mvect v_{i} = \mvect b$, where $\mvect b=-\mbf G\mvect v_{i-1}$, $ 1 \le i\le m$, in Eq. (\ref{eq_krylov}). 
$\mbf C$ may be singular, relatively complicated or denser than $\mbf G$ (Fig. \ref{fig_matrix_distribution}) \cite{Yu14_TCAD, Yu2013, Zhuang2012}. 
If $\mbf C$ is singular, we should apply regularization process to make a non-singular $\mbf C$. 
It is time-consuming and impractical for large designs \cite{Weng12_TCAD, Chen12_TCAD}. 


\subsection{Invert Krylov Subspace Method}

We adopt a technique called as \emph{invert Krylov subspace} \cite{Zhuang14_DAC} to avoid the matrix factorization of $\mbf C$.  
It also helps capture important eigenvalues for exponential matrix function and accelerate MEVP evaluation. 
In \cite{Zhuang14_DAC}, invert Krylov subspace method is the second on convergent rate and the length of step-size $h$ after rational Krylov subspace method. 
However, its basis generation can be cheaper for general nonlinear circuit simulation problems. 
Besides, the properties fit well with nonlinear dynamical  systems where the step-size is limited by the nonlinearity of the devices. 
Invert Krylov subspace is constructed as follows, 
\begin{eqnarray}
{\rm K_m}(\mbf  J^{-1} , \mvect  v) &=& \text{span}
{\{  \mvect v, \mbf  J^{-1}\mvect  v, \cdots, \mbf J^{-(m-1)} \mvect  v}\}  \\
&=& \text{span} \{ \mvect  v,  -\mbf G^{-1}\mbf C \mvect   v,   \cdots, 
(-\mbf G^{-1}\mbf C)^{(m-1)} \mvect  v \}.  \nonumber 
\end{eqnarray}
During the process, only $\mbf G$ is required for matrix factorization.
Therefore, we are able to gain computational advantages when $\mbf C$ contains large number of non-zeros to describe many parasitics and strongly coupling effects.
\begin{eqnarray}
\mbf  J^{-1} \mbf V_m = \mbf V_m \mbf H_m 
+   h_{m+1,m} \mvect  v_{m+1} \mvect  e^{\mathbf T}_m
\end{eqnarray}
{{MEVP}}  of  $\mbf x (t) = e^{t\mbf  J}\mbf v =  e^{-t\mbf C^{-1}\mbf G}\mvect v$ is computed as 
    \begin{eqnarray}
    \mbf x_m (t) = \lVert \mvect v \rVert \mbf V_m e^{t\mbf H_m^{-1}} \mvect e_1.
    \end{eqnarray}
{{Error estimation of MEVP}} 
is derived to determine dimension size of Krylov subspace for MEVP. We have the derivative of $\mvect x_m(t)$  
\begin{eqnarray}
     \frac{d\mvect x_m(t)}{dt} =  \lVert \mvect v  \rVert \mbf V_m \mbf H_m^{-1} e^{t \mbf  H_m^{-1}} \mvect e_1.   
\end{eqnarray}
The residual is the difference between $ \frac{d\mvect x_m(t)}{dt}$ and $\mbf J \mvect x_m(t)$. 
Then, we use the following residual to 
check Kirchhoff's current and voltage laws (KCL/KVL). 
\begin{eqnarray}
\label{eq_inv_kry_err}
\mbf r_m(t)  &=& \mbf C\left(\frac{d\mvect x_m(t)}{dt}  - \mbf  J \mvect x_m(t)\right)  \nonumber \\  &=& \lVert  \mvect v \rVert \mbf C\left(  \mbf V_m \mbf H_m^{-1} e^{t \mbf  H_m^{-1}} \mvect e_1 - \mbf  J   \mbf V_m   e^{t \mbf  H_m^{-1}} \mvect e_1 \right)  \nonumber \\
&=& -\lVert \mvect v \rVert  h_{m+1,m}\mbf G \mvect v_{m+1}  \mvect e^{\mathbf T} \mbf H_m^{-1}   e^{t \mbf  H_m^{-1}} \mvect e_1    
\end{eqnarray}
We design {\emph{MEVP\_IKS}} in {\bf Algorithm \ref{alg_mevp}} and embed error estimator above as a convergence criteria to terminate Arnoldi process.
Then, given the initial state vector $\mvect v$, time step size $h_k$ and error budget $\epsilon$,
$\phi_1 \mvect v$ and $\phi_2 \mvect v$ are computed, respectively, as $\phi^{(e)}_1  $  and $\phi^{(e)}_2 $, where 
\begin{eqnarray}
 \phi^{(e)}_1 (\mbf C_k, \mbf G_k, \mvect v, \epsilon, h_k)   =  \frac{1}{h_k  \mbf J_k} \mvect m_{evp}  - \frac{1}{h_k \mbf J_k}  \mvect v  \nonumber  
\end{eqnarray} 
 \begin{eqnarray}
\phi^{(e)}_2 (\mbf C_k, \mbf G_k, \mvect v, \epsilon, h_k)   =   \frac{1}{h^2_k \mbf J^2_k}  \mvect m_{evp} 
-  \frac{1}{h^2_k \mbf J^2_k} \mvect v - \frac{1}{h_k \mbf J_k}  \mvect v  
 \end{eqnarray}
 where $\mvect m_{vep}$ is obtained from ${\emph{MEVP\_IKS}}(\mbf C_k, \mbf G_k, \mvect v, \epsilon, h_k)$.  
Note the denominator $J_k$ always helps cancel out $\mbf C^{-1}$ when forming the vector $\mvect v$  during the whole simulation, and leaves only $\mbf G^{-1}$, which is factorized by LU decomposition once per step and reused. 
The estimated nonlinear truncation error 
is 
 \begin{eqnarray}
 \label{eqn_nl_err1}
 e^{(e)}_{rr}(\mvect x_{k+1}, \mvect x_k) = h_k \phi^{(e)}_1 (\mbf C_k, \mbf G_k, \mbf C^{-1}_k \Delta \mvect F_k, \epsilon, h_k). 
 \end{eqnarray}
 This requires another MEVP via Krylov subspace. 
 When $e^{(e)}_{rr}$ is larger than given error budget, we reduce the time step $h_k = \alpha h_k$ ($\alpha < 1$). 
 In Sec \ref{sec_nonlinear_correct}, we add a correction term using  $\phi^{(e)}_2$ 
 and computed $\Delta \mvect F_k$, then the corrected solution is 
 \begin{eqnarray}
 \mvect x_{k+1,c}  = \mvect x_{k+1} - \gamma h_k \phi^{(e)}_{2} \left(\mbf C_k, \mbf G_k, \mbf C^{-1}_k  \Delta \mvect F_k, \epsilon, h_k \right).
 \label{eq_erc}
 \end{eqnarray}
 

\begin{algorithm}[t]
\caption{\emph{MEVP\_IKS}: Arnoldi Algorithm of Invert Krylov Subspace Method for MEVP $e^{-h\mbf  C^{-1} \mbf G}\mvect v$}
\label{alg_mevp} 
 \small
\KwIn { $\mbf C, \mbf G,   \mvect v, \epsilon,  h, m_c$  }
\KwOut {$ \mvect m_{evp}, \mbf V_m, \mbf H_m, m$}
{   
$\mvect v_1 = \frac{\mvect v }{\lVert \mvect v \rVert}$\;
       	\For {$j=1 : m_{c}$}
           {
      	Solve $-\mbf G \mvect w = {\mbf C } \mvect v_j$ and obtain $\mvect w$\; \label{algo_cgv} 
       	\For {$i = 1:j$}
       	{
        	$h_{i,j} = \mvect w^T\mvect v_{i}$\;
       	    $ \mvect w = \mvect w - h_{i,j} \mvect v_{i}$\;
       	}
       	$h_{j+1,j} = \lVert \mvect w \rVert $\;
       	$\mvect v_{j+1} = \frac{\mvect w }{h_{j+1,j}} $\;
       	\If {$ || \mbf r_m(h) || < \epsilon$ {\em{using \text{Eq. (\ref{eq_inv_kry_err})}}    } } 
       	{
       	   $m = j$\;
       	   break\; 
       	} 
       }
        $ \mvect m_{evp}  = \lVert \mvect v \rVert \mbf V_m 
       e^{h\mbf H_m^{-1}} 
         \mvect e_1$ \;   
       }
\end{algorithm} 

\begin{algorithm}[t]
     \label{algo:mexph}
     \caption{{\bf ER} and {\bf ER-C} Methods: Circuit Simulation Using Exponential Rosenbrock-Euler Integrators}
     \small
     \KwIn{Circuit netlist; $E_{rr}$ is the error budget for circuit simulator, $\epsilon$ criteria for the convergence within MEVP\_IKS; Corrected option $O_{pt\_c}$ is enabled for {\bf ER-C}, otherwise it is \bf{ER}.}
         \KwOut{Voltage/current solution 
         for time period $[0, T]$}
     {   
     Initialization phase: (a) load the netlist;
      (b) build linear matrices.
      	Set $t=0, k = 0, h_k = h_{init}$\; 
     	$\mbf x(0) = \mbf x_k =$  DC\_solution\;  
         \While{$t\le T$}{
     	Derive  $\mbf C_k, \mbf G_k, \mvect F_k$, $ \mvect g_{k}, \mvect  b_{k}$ from device models at  $\mvect x_k$\;
     	$LU\_decompose \left(\mbf G_k \right)$\;  
     	\label{algo_lu}
     	Compute $\phi^{(e)}_1$, $\phi^{(e)}_2$ 
     	 using {\emph {MEVP\_IKS}} ({\bf Algorithm \ref{alg_mevp}}) with above matrices and factorized ones.
     	 Obtain 
     		\label{reuse_set}  
     	$\left( \mvect m_{evp}, \mbf H_m, \mbf V_m, \mvect v, m \right)$ for $\phi^{(e)}_1$ and $\phi^{(e)}_2$, respectively\;
     	Set $i=0$  \tcp*[l]{ \footnotesize the iteration index}
     	\While{true}{
     	    \label{algo_error_iteration_start}  
     	    Obtain $\mvect x_{k+1}$ by Eq. (\ref{eq_rbe}) with the given/updated  step size $h_k$ and previous sets of $\left( \mvect m_{evp}, \mbf H_m, \mbf V_m, \mvect v, m \right)$ from line \ref{reuse_set}\;
     	    Derive  $\mvect F_{k+1}$ from device models at  $\mvect x_{k+1}$ and obtain $\Delta \mvect F_{k} = \mvect F_{k+1}- \mvect F_k$\; 
     	    Compute $e^{(e)}_{rr}(\mvect x_{k+1}, \mvect x_k) = h_k  \phi^{(e)}_1  (\mbf C_k,\mbf G_{k},  \mbf C^{-1}_k \Delta \mvect F_{k+1}, \epsilon, h_k)$ by {\emph {MEVP\_IKS}}\;
     	  	\If{ $O_{pt\_c} $ is enabled}{
     	  	\label{algo_erc_start} 
      	 	$\mvect D= \gamma h_k  \phi^{(e)}_2( \mbf C_k,\mbf G_k, \mbf C^{-1}_k \Delta \mvect F_k, \epsilon, h_k)$ by {\emph {MEVP\_IKS}}\;
     	  	$\mvect x_{k+1} = \mvect x_{k+1} - \mvect D$
     	  	\tcp*[l]{{\footnotesize {ER-C, $e.g., \gamma = 0.1$}}}
     	  	\label{algo_erc_end}
     	  	}
     	    \If{  $\lVert e^{(e)}_{rr}(\mvect x_{k+1}, \mvect x_k) \rVert_{\infty}  \le E_{rr}$}{ 
     		$\mvect x(t+h) = \mvect x_{k+1}$\;
     		break\;
     	    }
     	     $i = i + 1$, $h = \alpha h$\tcp*[l]{update $h_k$, i.e., $\alpha=1/2$}
     	    \label{algo_error_iteration_end}
     	}
     	$t=t+h_k,~k=k+1$\;
     	\If{$i$ is small enough}{   \label{algo_fast}
     	    \tcc{e.g., update $h_k$ when $i < 5$} 
     	    $h_k = \beta h_k$\tcp*[l]{e.g., $\beta = 2$}
     	    }
         }
     }
     \label{algo_erik}
\end{algorithm}
 \vspace{-0.4cm}
\subsection{Overall Circuit Simulation Framework} 
{\bf {Algorithm \ref{algo_erik}}} shows our circuit simulation framework. 
The proposed method takes only one LU decomposition per time step $h$ (line \ref{algo_lu}), while BENR makes at least two LU decompositions (one for integration and NR's convergence check).
When the solutions are not converged, BENR updates the time step and repeats iterations including LU operations. 
Moreover, based on the explicit formulation, 
our method does not need to repeat LU decomposition when we adjust the length $h$ of time steps for error controls.  When the error is beyond the error budget $E_{rr}$,  our framework easily adjusts time step without extra LU operations to reduce the nonlinear error (from line \ref{algo_error_iteration_start} to \ref{algo_error_iteration_end}). 
Note that there is  no need to perform LU decomposition of $\mbf G$. 
There is no matrix factorization with $\mbf C$.
In addition, invert Krylov subspace method deals with a much simpler matrix $\mbf G$, while BENR method uses the format of $\left( \mbf G+ \frac{\mbf C}{h} \right)$. 
If $\mbf C$ contains more detailed  parasitics, 
BENR is affected more than our method (more non-zeros in $\mbf C$ indeed increase the  operations of sparse matrix and vector multiplication when we construct Krylov subspace in our framework).
If the option of correction term $O_{pt\_c}$ is enabled, line \ref{algo_erc_start} to \ref{algo_erc_end} will perform extra computations to provide more accurate solution, leading to {\bf ER-C}, the method of exponential Rosenbrock-Euler with correction term (i.e., Eq. (\ref{eq_ferc})).  
Without this option, the method is plain {\bf ER} (i.e., Eq. (\ref{eq_rbe})). 
Line \ref{algo_fast} shows that fast convergence rate triggers step-size increase.


\section{Results}
\label{sec_result}

{ 
The algorithms are implemented in MATLAB and C/C++, where all devices are evaluated by BSIM3.
The interactions between C/C++ and MATLAB2013a are through MATLAB Executable (MEX) external interface with GCC 4.7.3.  
We test our algorithms in a Linux workstation (Intel CPU i7 3.4GHZ and 32GB memory). 
All of algorithms and procedures utilize single thread mode. All of test cases in our experiment are stiff designs with singular $\mbf C$ matrices.  
Hence, we do not use previous matrix exponential method in \cite{Weng12_TCAD, Weng12_ICCAD},  
since it requires extra time-consuming regularization processes at each time step just to make $C$ to be non-singular.  
We compare our framework to the baseline circuit simulator using 
the standard
backward Euler with Newton-Raphson method (BENR). 

A stiff nonlinear circuit containing a inverter chain 
is used to demonstrate the favorable characteristics of our proposed method. We extract waveforms from one observed node of that circuit to compare the accuracy of BENR,  exponential Rosenbrock-Euler method (ER) and ER with  correction term (ER-C). In Fig. \ref{fig_inv_chain}, compared to the reference solution (REF) obtained by BENR with smaller step size ($10^{-14}s$), 
our ER and ER-C are more accurate than BENR.
\begin{figure}[h]
\vspace{-0.05in}
\begin{subfigure}[t]{0.21\textwidth}
\centering
\includegraphics[trim = 0mm 3mm 10mm 11mm, clip, keepaspectratio, width=1.5\textwidth]{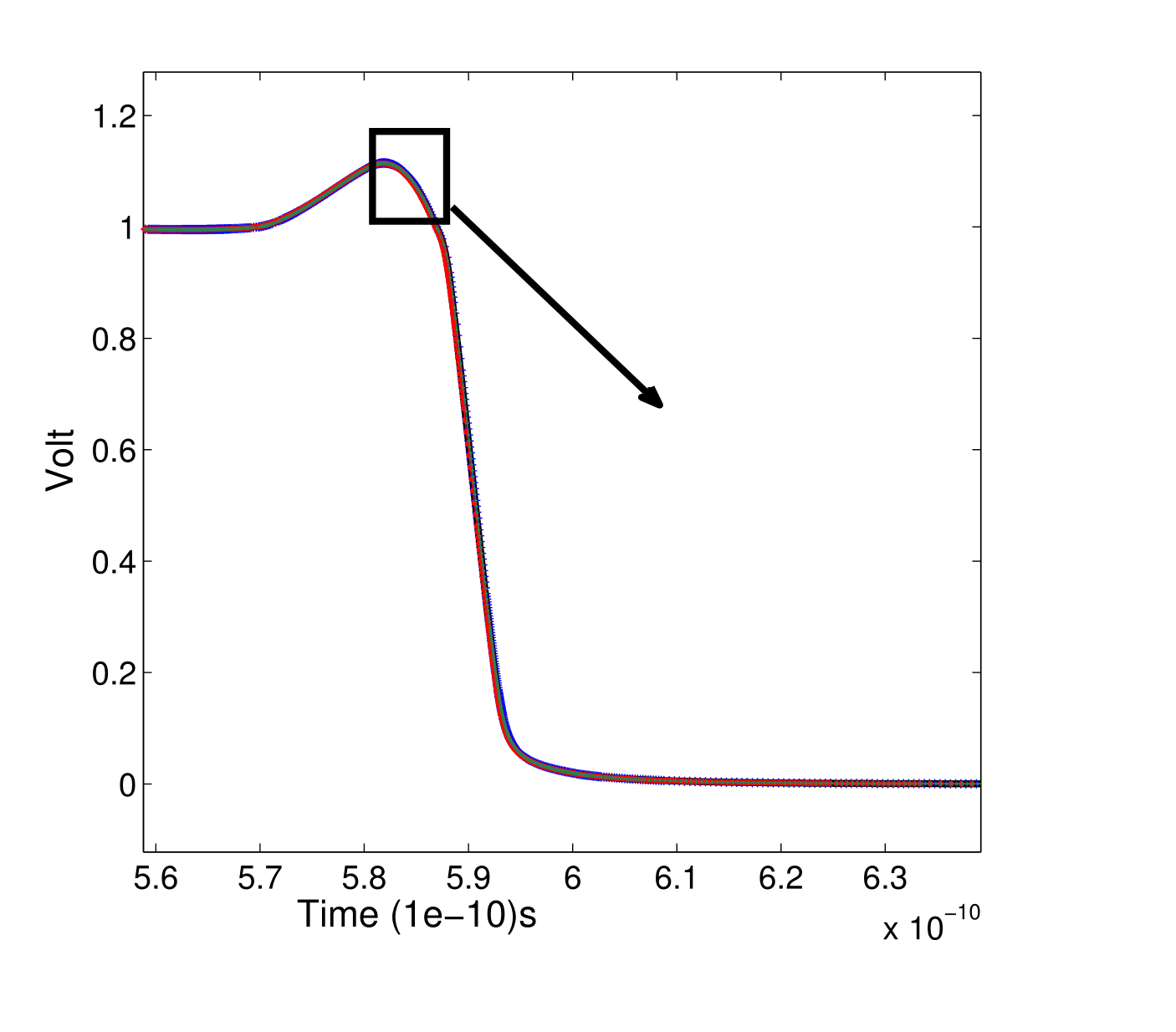}
\end{subfigure}
\begin{subfigure}[t]{0.23\textwidth}
\centering
\includegraphics[trim = 0mm 3mm 19mm 10mm, clip, keepaspectratio, width=1.1\textwidth]{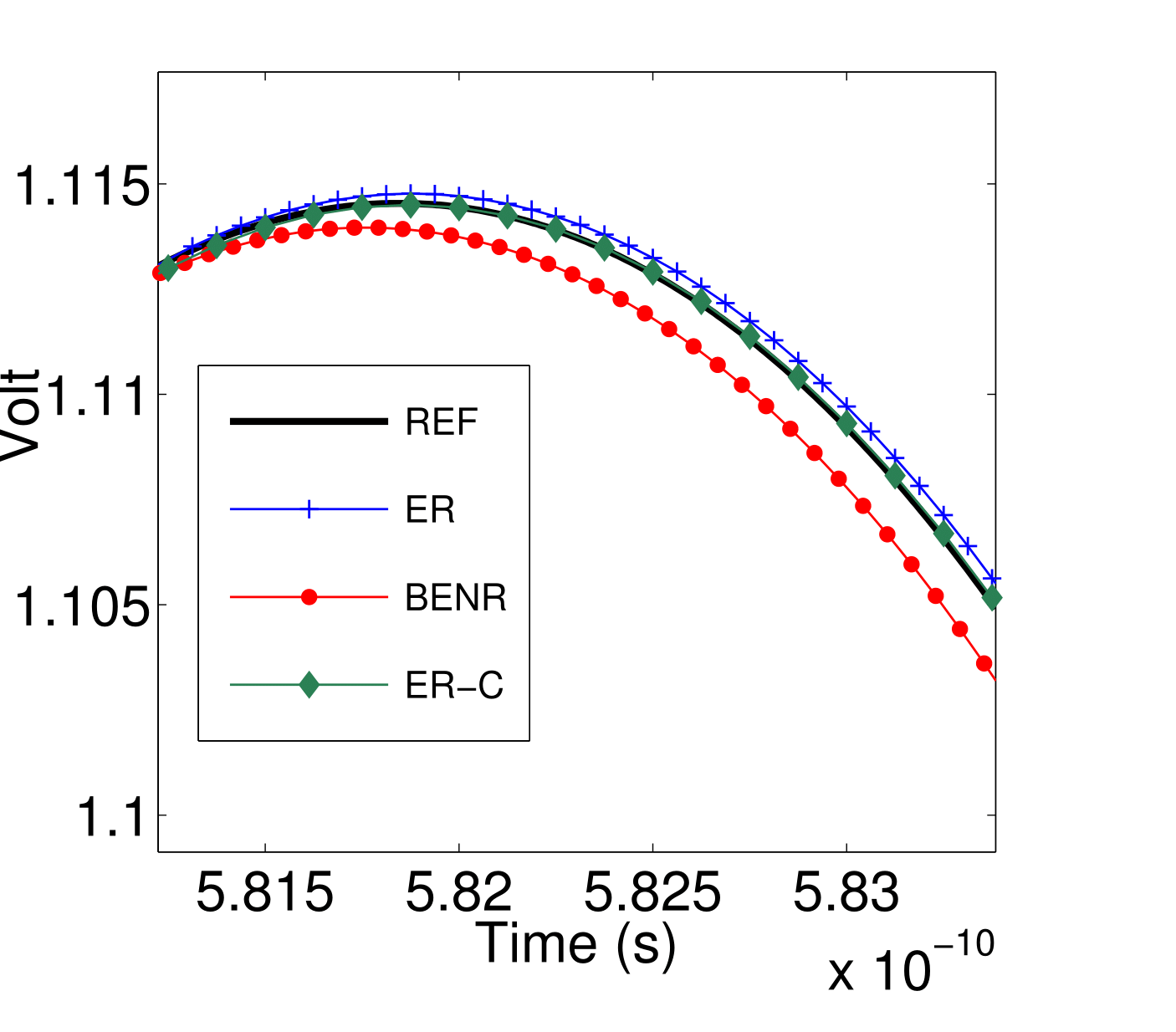} 
\end{subfigure}  
\vspace{-0.1in}
\caption{Accuracy comparisons of transient simulation solutions obtained by BENR, ER and ER-C. 
REF is the reference solution obtained from BENR with step size $10^{-14}s$. BENR and ER both use step size $10^{-13}s$. ER-C uses 2X step size as BENR and ER and still maintain better accuracy.} 
\label{fig_inv_chain} 
\end{figure}

In Table \ref{tab_result}, we list the relatively large test cases with their specifications. 
We compare the runtime performance and capability of those large test cases among BENR and our proposed algorithms ER, ER-C. 
We set $\epsilon=10^{-7}$ in {\bf Algorithm} \ref{alg_mevp} for MEVP convergence condition. 
We try test cases $ckt1$ to $ckt3$, which have capacitances matrices $\mbf C$ with extremely sparse non-zeros distributions.
Our framework  achieves speedups, that is 1.8$\times$ by ER and 1.4$\times$ by ER-C.
The small speedups are because of their relatively simple $\mbf C$ matrices. LU decomposition of $\left( \frac{\mbf C}{h}+\mbf G \right)$ is dominated by the part of $\mbf G$.
 ER and ER-C need LU decomposition of $\mbf G$ for invert  Krylov subspace constructions as well. Therefore, our framework is not fully beneficial here. 
However, we notice the reduced numbers of LU operations and time steps improve the runtime performance and compensate the portion of inverted Krylov subspace generations.

Cases from $ckt4$ to $ckt8$ are more challenging since they have more complicated capacitance matrices. 
For the available speedup numbers, we achieve speedups by magnitude, {\bf 23.2}$ \times$ by ER and {\bf 20.7}$\times$ by ER-C. 
For the speedup numbers not available, BENR cannot complete the simulation tasks under our limited computing resources in a reasonable time range. However, our framework processes those tasks efficiently.  
Note the design of $ckt5$ contains interconnect structure of FreeCPU (Fig. \ref{fig_matrix_distribution}), and there are corresponding 40 drivers similar to $ckt3$.
In such simulation with stronger parasitic couplings,   
BENR needs 54K seconds to finish the whole simulation. In contrast, the runtime performance of our methods is more stable and achieves over {\bf 35$\times$} speedups. 
The increasing runtime compared to our methods on $ckt3$ is partially due to the matrix and vector multiplication, e.g., $\mbf C \mvect v$ (line \ref{algo_cgv} in {\bf Algorithm} \ref{alg_mevp}) during the invert Krylov subspace generations since $C$ has larger $nnz$. 
$ckt4$ has more nonlinear MOSFET devices but simpler coupling capacitances than $ckt5$.
The over {\bf 4$\times$} speedups are still observed from our methods. 
Cases $ckt6$ to $ckt8$ are even more challenging ones with many parasitics in certain parts of $C$. 
LU decompositions in BENR exhaust our machine's memory (32GB).
Thus, we mark them  
as ``Out of Memory'' in Table \ref{tab_result}. {\bf{NA}} (no available) denotes the speedups number are not available, which represents our methods' capability of handling those challenging cases while standard BENR cannot. 
Our methods only need to factorize $\mbf G$. 
In these three cases, the peak memory consumption of our framework is smaller than 12GB reported by Linux $top$ command for corresponding process of MATLAB instance (BENR costs at least 32GB).   
Our methods can run through all the simulation tasks and the solutions converge to the version of ER-C with smaller step-sizes. 
The benchmark performance shows our algorithmic framework's advantages on memory usage and runtime performance. 

\begin{table*}[ht]
	\caption{Simulation Results.}
	{     {\bf \#N}: the number of unknowns;  
		{\bf \#Dev.}: the number of nonlinear devices. 
		{\bf nnz$_{\mbf C}$} and {\bf nnz$_{\mbf G}$}: the number of non-zero elements in linear capacitance/inductance matrix $\mbf C$ 
		and conductance/resistance matrix $\mbf G$.  
		{\bf \#step}: the number of steps for transient simulation;
		{\bf \#NR$_a$}: the average number of Newton-Raphson iterations for each time step.
		{\bf \#m$_a$}: the average dimension number of invert Krylov subspace for each time step. 
		{\bf RT(s)}: the runtime of transient simulation. 
		{\bf SP}: the runtime speedup of proposed method over BENR (backward Euler formulation with Newton-Raphson iterations). 
	}
	\begin{center}
		\small
		\label{tab_result}
		\begin{tabular}{|c|r|r|r|r||r|r|r||r|r|r|r||r|r|r|r|}
			\hline
			\multicolumn{5}{|c||}{Designs \& Specifications} 
			& \multicolumn{3}{|c||}{BENR} 
			& \multicolumn{4}{|c||}{ER} 
			& \multicolumn{4}{|c|}{ER-C}  
			\\ 
			\cline{1-16}
			Case & \#N   & \#Dev. & nnz$_C$ & nnz$_G$ 
			& \#step & \#NR$_{a}$ & RT(s)    
			& \#step  & \#m$_{a}$ & RT(s)  & SP  
			& \#step  & \#m$_{a}$ & RT(s)  & SP  
			\\ \hline  \hline
			
			$ckt1$ & 84K & 40K & 19K   &  188K
			& 1950  & 2.8  & 1983.6    
			& 1398 & 30.2 & 1314.6   &   1.5$\times$ 
			& 1404  &  31.6 & 1461.6 
			&  1.4$\times$ \\ \hline	
			$ckt2$ &  3.3M  & 0.1M    & 1.6M  & 9.8M 
			& 2375 & 3.1 & 122926.2 
			& 1995 & 28.4 & 78042.2 & 1.6$\times$
			& 2005 & 32.7 & 96406.3 & 1.3$\times$
			\\ 	\hline   	
			$ckt3$ & 54K   & 40 &  19K   &   181K &
			1928	& 2.8 & 1497.4
			&  1380	& 27.4  &      659.0 & 2.3$\times$    
			& 1418   &31.1   
			& 888.9  &      1.7$\times$
			\\ \hline                            
			$ckt4$  & 84K & 40K  & 37K 
			& 188K
			& 1950 & 2.8 & 6153.0  
			&   1372 & 27.7  &  1138.6 &  {5.4$\times$}  
			&  1422 & 30.3  & 1411.4 & {4.4$\times$} 
			\\ 	\hline 	
			$ckt5$  & 54K   & 40 & 82K  & 181K & 
			1917 & 2.8 & 53873.5 
			& 	1353	& 27.8   & 1298.2 & {41$\times$} 
			&    
			1365	& 31.5  & 1447.4  &  {37$\times$}  
			\\ 	\hline 
			$ckt6$  & 84K & 40K  &  157K & 188K  &  
			\multicolumn{3}{c||}{Out of Memory}   
			& 1282 &  27.0 & 1447.0   &  {\bf  NA} 
			& 1299 & 31.2 
			& 1917.7  
			&   {\bf  NA} 
			\\ 	\hline       
			$ckt7$ &  0.2M  & 0.1M   
			& 0.2M & 0.6M &
			\multicolumn{3}{c||}{Out of Memory} 
			& 1395  & 26.7  &   8016.6   &  {\bf  NA} 
			&  1399 & 30.4  &  9373.8 & {\bf  NA} 
			\\ 	\hline  
			$ckt8$ &  3.3M  & 0.1M    & 1.7M  & 9.8M &  
			\multicolumn{3}{c||}{Out of Memory}   
			& 2310  & 29.3 & 86386.7   &   {\bf  NA} 
			&  2360 & 34.2  & 98681.7   &    {\bf  NA} 
			\\ \hline
		\end{tabular}
	\end{center}
	\vspace{-0.4cm}
\end{table*}

\section{Conclusion}
\label{sec_conclusion}
We propose a new and efficient algorithmic framework for time-domain large-scale circuit simulation using exponential integrators. We also devise a correction term on top of this formulation and further improve the accuracy. 
By virtue of the stable explicit nature of our formulation, we remove  Newton-Raphson iterations and reduce the number of LU decomposition operations. 
In this framework, 
matrix exponential and vector product is computed by efficient invert Krylov subspace. This approach can keep capacitance/inductance matrix from matrix factorization and avoid the time-consuming regularization process when there are singular capacitance/inductance matrices during the simulation. Moreover, this method does not need to repeat LU decompositions when we adjust the length of time steps for error controls.  
Compared to conventional methods, our new framework has several distinguished features. The proposed method can handle many parasitics, strongly coupled or post-layout circuits, even when conventional methods are not applicable.
We test our proposed framework against standard backward Euler method with Newton-Raphson iterations, called as BENR. Our framework does not only accelerate the simulation, but also manages to finish the challenging test cases when BENR processes extremely slow or even fails. 
Many aspects are worthwhile to investigate within this new circuit simulation framework. For instance, matrix and vector multiplication plays an important role in our framework. 
Parallel processing this kind of operation by multi-core/many-core architectures could be beneficial to enhance the runtime performance. 

\section*{Acknowledgment}
We acknowledge the support from NSF CCF-1017864. Hao Zhuang is supported by  the UCSD Powell Fellowship and Qualcomm FMA Fellowship. Xinan Wang is supported by the UCSD Jacobs Fellowship.
This work is also partially supported by NSFC (grant No. 61422402).
We thanks 
Prof. Mike Botchev, Dr. Quan Chen, Mr. Ryan Coutts, Prof. Marlis Hochbruck, 
Mr. Chia-Hung Liu, Dr. John Loffeld, Dr. Jingwei Lu,  
Prof. Alexander Ostermann,  
Prof. Mayya Tokman, Dr. Lining Zhang and Mr. Xiang Zhang for the helpful discussions.

}

{ 
\bibliographystyle{ieeetr}
 \footnotesize{
\bibliography{CADRefs} 
}
}

\clearpage
\end{document}